# Fractional quantum Hall effect and insulating phase of Dirac electrons in graphene

Xu Du[1]†, Ivan Skachko[1], Fabian Duerr[1], Adina Luican[1] & Eva Y. Andrei[1]

[1]Department of Physics and Astronomy, Rutgers University, Piscataway, New Jersey 08855, USA.

†Present address: Department of Physics and Astronomy, Stony Brook University, Stony Brook, New York 11794-3800, USA.

**In graphene, which is an atomic layer of crystalline carbon, two of the distinguishing properties of the material are the charge carriers' two-dimensional and relativistic character. The first experimental evidence of the two-dimensional nature of graphene came from the observation of a sequence of plateaus in measurements of its transport properties in the presence of an applied magnetic field[1,2]. These are signatures of the so-called integer quantum Hall effect. However, as a consequence of the relativistic character of the charge carriers, the integer quantum Hall effect observed in graphene is qualitatively different from its semiconductor analogue[3]. As a third distinguishing feature of graphene, it has been conjectured that interactions and correlations should be important in this material, but surprisingly, evidence of collective behaviour in graphene is lacking. In particular, the quintessential collective quantum behaviour in two dimensions, the fractional quantum Hall effect (FQHE), has so far resisted observation in graphene despite intense efforts and theoretical predictions of its existence[4–9]. Here we report the observation of the FQHE in graphene. Our observations are made possible by using suspended graphene devices probed by two-terminal charge transport measurements[10]. This allows us to isolate the sample from substrate-induced perturbations that usually obscure the effects of interactions in this system and to avoid effects of finite geometry. At low carrier density, we find a field-induced transition to an insulator that competes with the FQHE, allowing its observation only in the highest quality samples. We believe that these results will open the door to the physics of FQHE and other collective behaviour in graphene.**

The description of graphene in terms of a two-dimensional (2D) zero-bandgap semiconductor with low energy excitations represented by non-interacting Dirac fermions is surprisingly successful[3]. Indeed, most experimental results thus far are captured by this single particle picture, in which collective effects and interactions are assumed to be negligibly small. In particular, scanning tunnelling spectroscopy in a transverse magnetic field, $B$, revealed a sequence of Landau levels with energy $E_n = \text{sign}(n)\sqrt{2e\hbar v_F^2 |n| B}$, providing the most direct evidence of the non-interacting Dirac fermion picture[11]. (Here $n = 0, \pm 1, \pm 2$ is the Landau level index, $e$ is the elementary charge, $\hbar = h/2\pi$ where $h$ is Planck's constant and $v_F$ is the Fermi velocity[3]) Sweeping the field or carrier density, $n_s$, ( the subscript s, distinguishes the carrier density from the integers n designating the Landau level sequence) through these Landau levels in a magneto-transport measurement reveals quantum-Hall conductance plateaus (the integer quantum Hall effect) at values $G_{xy} = \nu \frac{e^2}{h}$ for filling factors $\nu = \frac{n_s h}{B e} = 4(n+1/2)$ where all available states up to the n'th Landau level are occupied. . Here 4 is due to the spin and valley degeneracy[3] and the $\pm$ signs reflect the electron–hole symmetry. The $\frac{1}{2}\frac{4e^2}{h}$ offset, absent in non-relativistic 2D electron systems (2DES), is a result of the special status of the $n = 0$ Landau level for the massless Dirac fermions: half of its states are hole states, and the other half are electron states. This picture is expected to fail when interactions lift the degeneracy, resulting in new integer plateaus outside this sequence[12-15]. Furthermore, strong correlations between the electrons are expected to give rise to plateaus at fractional filling factors, reflecting the condensation into new ground states[4-9].

Thus far, magneto-transport experiments on non-suspended graphene samples show no evidence of interactions or correlations for fields below ~25 T. In higher fields[16], the appearance of quantum Hall effect plateaus at $\nu = 0, \pm 1, \pm 4$ suggests that interaction effects do exist in graphene, but only become observable when their energy scale exceeds that of the fluctuations due to random potentials induced by external sources. Similarly, the insulating state at $\nu = 0$ observed in some non-suspended samples in strong magnetic fields[17] but not in others[18] suggests that extrinsic effects play an important role in obscuring the underlying intrinsic physics of the charge carriers in graphene. Therefore in order to understand the role of correlations in the low density phases and to solve the long-standing mystery of whether graphene can support an FQHE, it is necessary to better control sample quality.

Recently, a significant improvement in transport properties was demonstrated in suspended graphene samples where substrate-induced perturbations were eliminated[10,19]. The combination of ballistic transport and low carrier density achieved in suspended graphene is particularly well suited for studying the intrinsic properties of this system. To ensure mechanical and structural integrity of the sample, the suspended devices are quite small, with typical dimensions of length $L \approx 0.5$–1 μm and width $W \approx 1.5$–3 μm. Surprisingly, in these small devices the standard Hall-bar measurement geometry fails to yield the expected quantum Hall effect features[19]. A possible cause, which we discuss in detail elsewhere[29], is that the proximity between voltage and current leads in such small samples shorts out most of the Hall voltage. This is a consequence of the peculiar potential distribution at large Hall angles (the case of plateaus in the quantum Hall effect) where most of the potential drop, roughly equal to the Hall voltage, occurs at opposite corners of the sample close to the current leads[20], also known as hot spots in the Hall effect literature. A metallic lead placed within this region necessarily shorts out the Hall voltage. As suspended graphene devices (typically of micrometre size) are too small to allow the voltage leads to be placed outside the hot spot regions, shorting the Hall voltage is almost unavoidable[19]. This problem is circumvented in the two-terminal lead configuration[10] used in the present work (Fig. 1a). We note that the classical contact resistance of the two-terminal devices discussed here is negligible compared to the quantum resistance, as indicated by the very small deviation of the two-terminal resistance at the quantum Hall effect plateaus from the standard values, described below.

All the suspended graphene samples studied here are in the ballistic regime, as measured by the density dependence of the zero field conductivity. In the hole-carrier sector, we find the mean-free-path $l_{\text{mfp}} \approx L/2$ and the conductivity $\sigma \propto n_s^{1/2}$, as expected for ballistic transport (Fig. 1b, c). Furthermore the lowest carrier density, typically $n_{s0} \approx (2$–$10) \times 10^9$ cm$^{-2}$, is more than an order of magnitude below that achieved in non-suspended samples, attesting to a much smaller density inhomogeneity[10]. For non-ballistic samples (graphene as well as 2DES in semiconductors), the sample quality is usually characterized by the carrier mobility. In ballistic graphene samples however, the value of mobility is meaningful only when it is associated with the carrier density at which it is measured. For the sample studied here, the Drude mobility, $\mu_D = \frac{\sigma}{n_s e}$, at $n_s \approx 10^{10}$ cm$^{-2}$ is 260,000 cm$^2$ V$^{-1}$ s$^{-1}$, and exhibits the $\propto n_s^{-1/2}$ dependence on carrier density expected for ballistic junctions (the field effect mobility at the same density is $\mu_{fe} = \frac{1}{e}\frac{d\sigma}{dn_s} \approx 200,000$ cm$^2$ V$^{-1}$ s$^{-1}$).

We studied the two-terminal magneto-transport in suspended graphene samples at temperatures ranging from 1.2 K to 80 K and fields up to 12 T. The relation between magneto-resistance oscillations and the quantum Hall effect measured in two-terminal devices is now well understood. It has been shown theoretically[21] that, for clean samples and low temperatures, the two-terminal conductance displays plateaus at values $G = \nu \frac{e^2}{h}$ that are precisely the same as the quantum Hall effect plateaus in the Hall conductance. In between the plateaus the conductance is non-monotonic, depending on the sample aspect ratio, $W/L$. In our devices where $W > L$, the conductance is expected to overshoot between plateaus, as is indeed observed (Fig. 1d). Our two-terminal measurements reveal well-defined plateaus associated with the anomalous quantum Hall effect that appear already in fields below 1 T. Above 2 T additional plateaus develop at $\nu = -1$ and at $\nu = 3$, reflecting interaction-induced lifting of the spin and valley degeneracy (Figs 2a and 3c). At low temperatures and above 2 T, we observe a FQHE plateau at $\nu = -1/3$ which becomes better defined with increasing field (Fig. 2a). When plotting $G$ versus $\nu$, the curves for all values of $B$ collapse together (Fig. 2b), and the plateaus at $\nu = -1/3, -1$ and $-2$ show accurate values of the quantum Hall conductance.

The FQHE in semiconductor based 2DES reflects the formation of an incompressible condensate, which can be described by a Laughlin wavefunction[22]. In the composite-fermion generalization of the FQHE[4,23], a strongly correlated electron liquid in a magnetic field can minimize its energy when the filling factor belongs to the series $\nu = \frac{p}{2sp \pm 1}$ (with $s$ and $p$ integers) by forming weakly interacting composite particles consisting of an electron and an even number of captured magnetic flux lines. In this picture, the FQHE with $\nu = 1/3$ corresponds to the integer quantum Hall effect with $\nu = 1$ for the composite particles consisting of one electron and two flux lines. Excitations out of this state would produce fractionally charged quasiparticles $q^* = e/3$, at an energy cost of the excitation gap, $\Delta_{1/3}$, which provides a measure of the state's robustness. It is not obvious a priori that the correlated state leading to the FQHE for the relativistic charge carriers in graphene is the same as that for the 2DES in semiconductors. In fact, several competing mechanisms have been discussed in the theoretical literature[4-9], involving states that break SU(4) symmetry as well as possible compressible, composite fermion Fermi sea states[7]. Interestingly, despite the qualitative difference in Landau level spectra between Dirac fermions in graphene and the non-relativistic electrons in semiconductors, the $\nu = 1/3$ state is formally expected to be the same in both cases[4,5] but with the pseudospin in graphene playing the role of the traditional electron spin in the non-relativistic case. In order to distinguish experimentally between the various mechanisms, it is useful to study the quasiparticle excitation energy. In multi-lead transport measurements, such as the Hall bar configuration, this can be obtained from the temperature dependence of the longitudinal conductance. However, in a two-terminal measurement it is not possible to separate the longitudinal and transverse components of the conductance. Nevertheless, an order of magnitude estimate can be obtained from the temperature at which the $\nu = 1/3$ plateau disappears. In Fig. 2c we note that this plateau smears out with increasing temperature and disappears above 20 K, suggesting that $\Delta_{1/3} \approx 20\,\text{K} \approx 0.008 E_C(12\text{T})$, where $E_C = \dfrac{e^2}{4\pi\varepsilon_0 \varepsilon l_B}$ is the Coulomb energy, $\varepsilon_0$ the permittivity of free space, $\varepsilon = 1$ the dielectric constant of the host material (vacuum) and $l_B = \sqrt{\dfrac{\hbar}{eB}}$ is the magnetic length. The discrepancy with the theoretical prediction, $\Delta_{1/3} \approx 0.1 E_C$, in a Laughlin-like condensate[4,5] is comparable to that seen in 2DES in semiconductors. There it is attributed to deviations from an ideal 2D system due to the finite thickness of the quantum wells (10–30 nm), disorder and mixing with higher Landau levels[24] Importantly, the value of $\Delta_{1/3}$ in suspended graphene is more than an order of magnitude larger than the corresponding gap in the 2DES in semiconductors[25] because of the lower dielectric constant ($\varepsilon = 1$ in suspended graphene compared to $\varepsilon \sim 12.9$ in GaAs/GaAlAs heterostructures).

Next we discuss transport near the Dirac point ($\nu = 0$). Models for lifting of the fourfold spin and valley degeneracy fall in two categories depending on whether the spin degeneracy is lifted first, producing a so called quantum Hall ferromagnet[12-14] or the valley degeneracy is lifted first which gives rise to magnetic catalysis[14,15]. Both cases predict insulating bulk, but the former supports counter-propagating edge states and thus is a conductor, whereas the latter with no edge states is an insulator. In the quantum Hall ferromagnet scenario, where both spin and valley degeneracy can be lifted for all Landau levels, plateaus at all integer $\nu$ are allowed. In contrast, the magnetic catalysis scenario does not permit plateaus at odd filling-factors other than $\nu = \pm 1$. Experiments addressing this issue in non-suspended graphene are inconclusive[16-18]. While tilted field experiments support the quantum Hall ferromagnet situation[16], the absence of clear plateaus at $\pm 3, \pm 5$ is consistent with magnetic catalysis. The fact that both insulating and conducting behaviour were reported further contributes to the uncertainty.

To address this question in suspended graphene samples, we studied four samples in fields up to 12 T and at temperatures ranging from 1 K to 80 K (Supplementary Information). All samples were insulating at $\nu = 0$ for high fields and low temperature. Consistently we found that the higher the sample quality, as measured by the residual carrier density, the sharper the transition and the earlier its onset (lower fields and higher temperatures). In our highest quality sample, the onset of insulating behaviour scales linearly with field. This is clearly seen in Fig. 3a, where the sharp onset of insulating behaviour at $|\nu| \approx 0.1$ is marked by a dramatic increase in resistance. In the best samples, the maximum resistance value is instrument-limited to ~1 GΩ. In lower quality samples, the insulating region is broader, the onset less sharp and the maximum resistance lower. Interestingly, the FQHE state was only observed in samples with narrow insulating regions, suggesting a competition between the two ground states. This is illustrated in Fig. 3b, where the insulating phase, having become broader after contamination, 'swallowed' the 1/3 plateau. Current annealing the sample brought it back almost to its pristine condition again, revealing the 1/3 plateau.

Can the suspended graphene data shed light on the nature of the insulating phase? The appearance of a plateau at $\nu = 3$, shown in Fig. 3c, favours the quantum Hall ferromagnet over the magnetic catalysis. However, since the quantum Hall ferromagnet supports counter-propagating edge states, this scenario is inconsistent with insulating behaviour at $\nu = 0$. A possible solution would entail a gap opening in the edge states and thus a mechanism to admix them. This would require a mechanism to flip spins and valleys, such as magnetic impurities or segments of zigzag edges[26]. An alternative explanation is that the system undergoes a transition to a new broken symmetry phase, such as a Wigner crystal or a more exotic skyrme phase[27,28]. In this case pinning would naturally lead to insulating behaviour.

To better understand the insulating phase, we studied the temperature dependence of the $v = 0$ state. The details of the temperature dependence of the maximum resistance ($R_{max}$) show strong sample-to-sample variation, but all curves fit a generalized activated form: $R_{max} = R_0 \exp(-T_0/T)^\alpha$ with $\alpha \approx 1/3 - 1$. For the sample in Fig. 3c, $\alpha \approx 1/2$ for all fields, with $T_0 \propto B^2$. This may provide a hint to the nature of the insulating state, but more work is needed to resolve this question.

In summary, the experiments described here demonstrate that Dirac electrons do exhibit strong collective behaviour leading to an FQHE, which becomes apparent in suspended samples probed with a two-terminal lead geometry, where the system is isolated from external perturbations. We find that the FQHE is quite robust, appearing at low temperatures in fields as low as 2 T and persisting up to 20 K in a field of 12 T. The effect is significantly more robust than in the semiconductor-based 2DES, reflecting the stronger Coulomb interaction and the more 2D nature of the 2DES in graphene. We further show that the FQHE state competes with an insulating phase centred at $v = 0$ that broadens in the presence of disorder and can destroy it. This may explain why, despite the large energy scale of the Coulomb interactions, the FQHE has until now resisted observation in graphene. The observation of the FQHE plateau at $v = -1/3$ demonstrates that the FQHE is a stable ground state for the 2D Dirac fermions in graphene, and that it is a distinctly different phase from the insulating state at $v = 0$. These findings pave the way to future studies of FQHE physics in the Dirac fermion system for the $n = 0$ Landau level as well for higher Landau levels, where new correlated states, unique to relativistic charge carriers, are expected to emerge.

References

1. Novoselov, K. S. et al. Two-dimensional gas of massless Dirac fermions in graphene. *Nature* **438**, 197–200 (2005).
2. Zhang, Y. B., Tan, Y. W., Stormer, H. L. & Kim, P. Experimental observation of the quantum Hall effect and Berry's phase in graphene. *Nature* **438**, 201–204 (2005).

3. Castro Neto, A. H., Guinea, F. & Peres, N. M. R., Novoselov, K. S &. Geim. A. K.   The electronic properties of graphene. *Rev. Mod. Phys.* **81**, 109-162 (2009).

4. Toke, C., Lammert, P. E., Jain, J. K. & Crespi, V. H. Fractional quantum Hall effect in graphene. *Phys. Rev. B* **74**, 235417 (2006).

5. Yang, K., Das Sarma, S. & MacDonald, A. H. Collective modes and skyrmion excitations in graphene SU(4) quantum Hall ferromagnets. *Phys. Rev. B* **74**, 075423 (2006).

6. Peres, N. M. R., Guinea, F. & Castro Neto, A. H. Electronic properties of disordered two-dimensional carbon. *Phys. Rev. B* **73**, 125411 (2006).

7. Khveshchenko, D. V. Composite Dirac fermions in graphene. *Phys. Rev. B* **75**, 153405 (2007).

8. Shibata, N. & Nomura, K. Coupled charge and valley excitations in graphene quantum Hall ferromagnets. *Phys. Rev. B* **77**, 235426 (2008).

9. Goerbig, M. O. & Regnault, N. Analysis of a SU(4) generalization of Halperin's wave function as an approach towards a SU(4) fractional quantum Hall effect in graphene sheets. *Phys. Rev. B* **75**, 241405 (2007).

10. Du, X., Skachko, I., Barker, A. & Andrei, E. Y. Approaching ballistic transport in suspended graphene. http://arXiv.org/abs/0802.2933 (2008);.*Nature Nanotechnol.* **3**, 491 (2008).

11. Li, G., Luican, A. & Andrei, E. Y. Scanning tunnelling spectroscopy of graphene. *Phys. Rev. Lett.* **102**, 176804 (2009)

12. Nomura, K. & MacDonald, A. H. Quantum Hall ferromagnetism in graphene. *Phys. Rev. Lett.* **96**, 256602 (2006).

13. Alicea, J. & Fisher, M. P. A. Integer quantum Hall effect in the ferromagnetic and paramagnetic regimes. *Phys. Rev. B* **74**, 075422 (2006).

14. Yang, K. Spontaneous symmetry breaking and quantum Hall effect in graphene. *Solid State Commun*. v.**143**, p. 27-32  (2007).

15. Gorbar, E. V., Gusynin, V. P., Miransky, V. A. & Shovkovy, I. A. Dynamics in the quantum Hall effect and the phase diagram of graphene. *Phys. Rev. B* **78**, 085437 (2008).

16. Zhang, Y. et al. Landau level splitting in graphene in high magnetic fields. *Phys. Rev. Lett.* **96,** 136806 (2006).

17. Checkelsky, J. G., Li, L. & Ong, N. P. Divergent resistance at the Dirac point in graphene: evidence for a transition in high magnetic field. *Phys. Rev. Lett*. **100**, 206801 (2008).

18. Abanin, D. A. et al. Dissipative quantum Hall effect in graphene near the Dirac point. *Phys. Rev. Lett.* **98**, 196806 (2007).

19. Bolotin, K. I. et al. Ultrahigh electron mobility in suspended graphene. *Solid State Commun.* **146**, 351–355 (2008).

20. Wakabayashi, J. & Kawaji, S. Hall effect in silicon MOS inversion layers under strong magnetic fields. *J. Phys. Soc. Jpn* **44**, 1839-1849 (1978).

21. Abanin, D. A. & Levitov, L. S. Conformal invariance and shape-dependent conductance of graphene samples. *Phys. Rev. B* **78**, 035416 (2008).

22. Laughlin, R. B. Anomalous quantum Hall effect: an incompressible quantum fluid with fractionally charged excitations. *Phys. Rev. Lett.* **50**, 1395-1398 (1983).

23. Jain, J. K. Composite fermion approach for the fractional quantum Hall effect. *Phys. Rev. Lett.* **63**, 199-202 (1989).

24. Xin, W. et al. Mobility gap in fractional quantum Hall liquids: effects of disorder and layer thickness. *Phys. Rev. B* **72**, 075325 (2005).

25. Boebinger, G. S. et al. Activation energies and localization in the fractional quantum Hall effect. *Phys. Rev. B* **36**, 7919-7929 (1987).

26. Shimshoni, E., Fertig, H. A. & Pai, G. V. Onset of an insulating zero-plateau quantum hall state in graphene. *Phys. Rev. Lett.* **102**, 206408 (2009).

27. Cote, R., Jobidon, J. F. & Fertig, H. A. Skyrme and Wigner crystals in graphene. *Phys. Rev. B* **78**, 085309 (2008).

28. Poplavskyy, O., Goerbig, M. O. & Morais Smith, C. Local density of states of electron-crystal phases in graphene in the quantum Hall regime. Preprint at http://arXiv.org/abs/09102.2518 (2009).

29. Skachko I., Du X., Duerr F., Luican A.,. Abanin, D. A, Levitov, L.S. & Andrei E. Y. Integer and Fractional Quantum Hall Effect in Two-Terminal Measurements on Suspended Graphene. Preprint at http://arXiv.org/abs/0902.1902 (2009)

**Acknowledgements** Work was supported by DOE DE-FG02-99ER45742, ICAM, Alcatel-Lucent and NSF-DMR-045673. We thank J. Jain, D. Abanin, L. Levitov, A. Akhmerov, V. Falko and H. Fertig, for discussions.

**Figure 1 Characteristics of the suspended graphene devices. a**, False-colour scanning electron microscopy image of a typical suspended graphene device. The two centre pads are used for both current and voltage leads, while the outer pads are for structural support. The lead separation is $L = 0.7$ μm, and the typical graphene width is 1.5–3 μm. **b**, Carrier density dependence of the resistivity of a suspended graphene device in zero field. The sharp gate control of resistivity near the Dirac point indicates a low level of perturbation from random potentials. **c**, Carrier density dependence of the mean free path, $l_{\mathrm{mfp}} = \dfrac{\sigma h}{2e^2 (\pi n_s)^{1/2}}$, of the sample in **b**. Note that on the hole branch, $l_{\mathrm{mfp}} \approx L/2$, as expected for ballistic junctions. **d**, Conductance of the suspended graphene sample as a function of filling factor $\nu$ for $B = 1$ T and $T = 1.2$ K. The plateaus seen at integer filling factors correspond to the quantum Hall effect, as discussed in the text. The maxima in between the plateaus agree with the theoretical expectations[21] for a two-terminal graphene junction with the geometry of our sample, $W/L > 1$. The quantum Hall plateaus are better defined and narrower for the hole branch (negative filling factors), indicating less scattering of hole carriers, consistent with the lower resistance and longer mean free path on the hole branch, as shown in **b** and **c**.

**Figure 2 FQHE in suspended graphene. a**, Gate voltage dependence of resistance for the sample in Fig. 1, at indicated magnetic fields and $T = 1.2$ K. Already at 2 T we note the appearance of quantum Hall plateaus outside the non-interacting sequence, with $R = \dfrac{1}{\nu}\dfrac{h}{e^2}, \nu = 1, 1/3$. **b**, Hole conductance as a function of filling factors for $B = 2, 5, 8, 12$ T, at $T = 1.2$ K, showing that the data for all fields collapse together. Quantum Hall plateaus with conductance values $G = v\frac{e^2}{h}, v = 1, 1/3$, appear at the correct filling factors of $v = -1, -1/3$. **c**, Temperature dependence of the quantum Hall plateau features. The plateaus at $v = -1/3, -1$ become smeared out with increasing $T$ and disappear for $T > 20$ K.

**Figure 3 Insulating behaviour at $v = 0$. a**, Resistance as a function of filling factor for magnetic fields $B = 1,2,3,4,5,6,8,10,12$ T. For $|v| < 0.1$, the resistance increases sharply with increasing magnetic field. The maximum resistance value measured above 8 T is instrument-limited. **b**, Competition between FQHE and insulating behaviour. The sample in Fig. 1 was warmed up to room temperature and re-cooled to 1.2 K. Owing to the condensation of contaminants on the graphene channel, the insulating regime became broader, swallowing the FQHE plateau at $v = -1/3$. On current annealing, the sample was re-cleaned almost to its pristine condition, causing the insulating regime to recede and the plateau at $v = -1/3$ to reappear. **c**, Quantum Hall effect plateaus of a suspended graphene sample which showed $v = 3$. **d**, Logarithmic plot of maximum resistance for $v = 0$ as a function of $T^{-1/2}$ for the field values shown in **a**. The solid lines are guides to the eye.

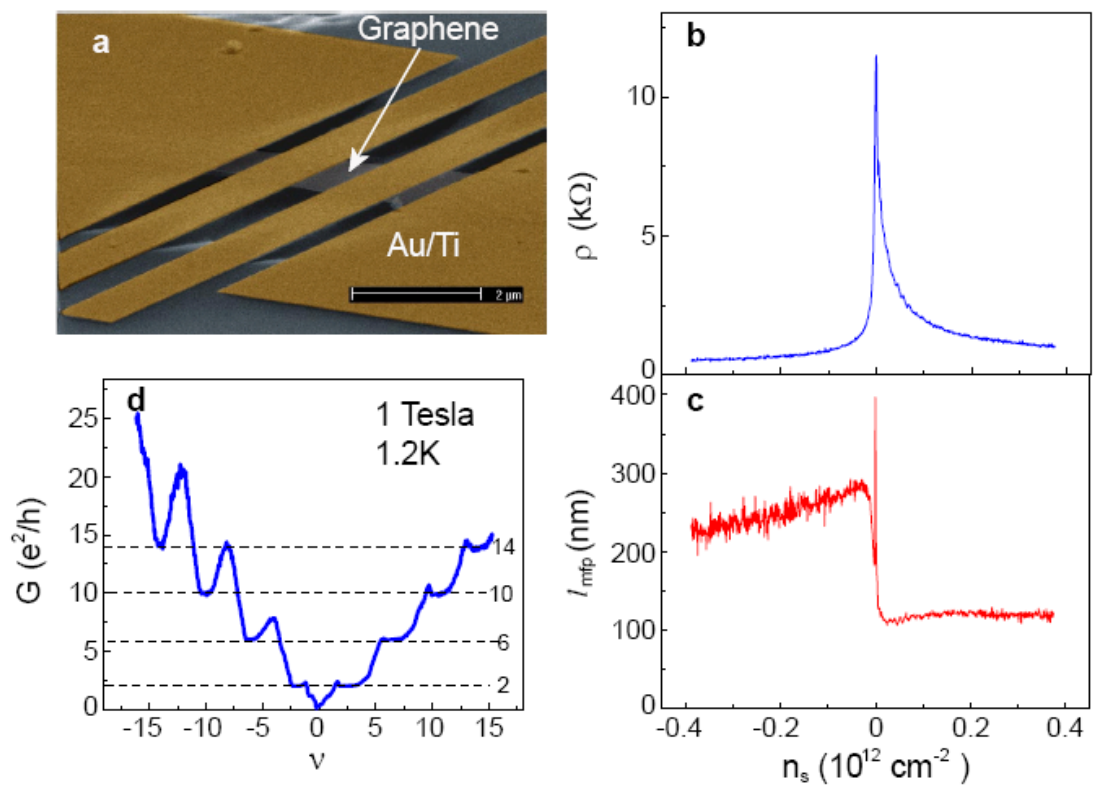

Figure 1

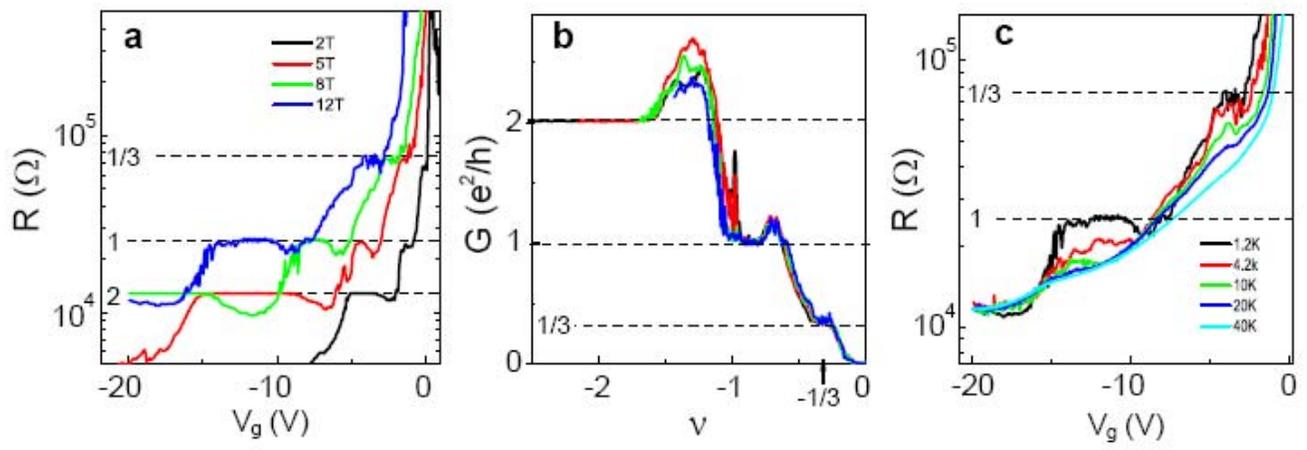

Figure 2

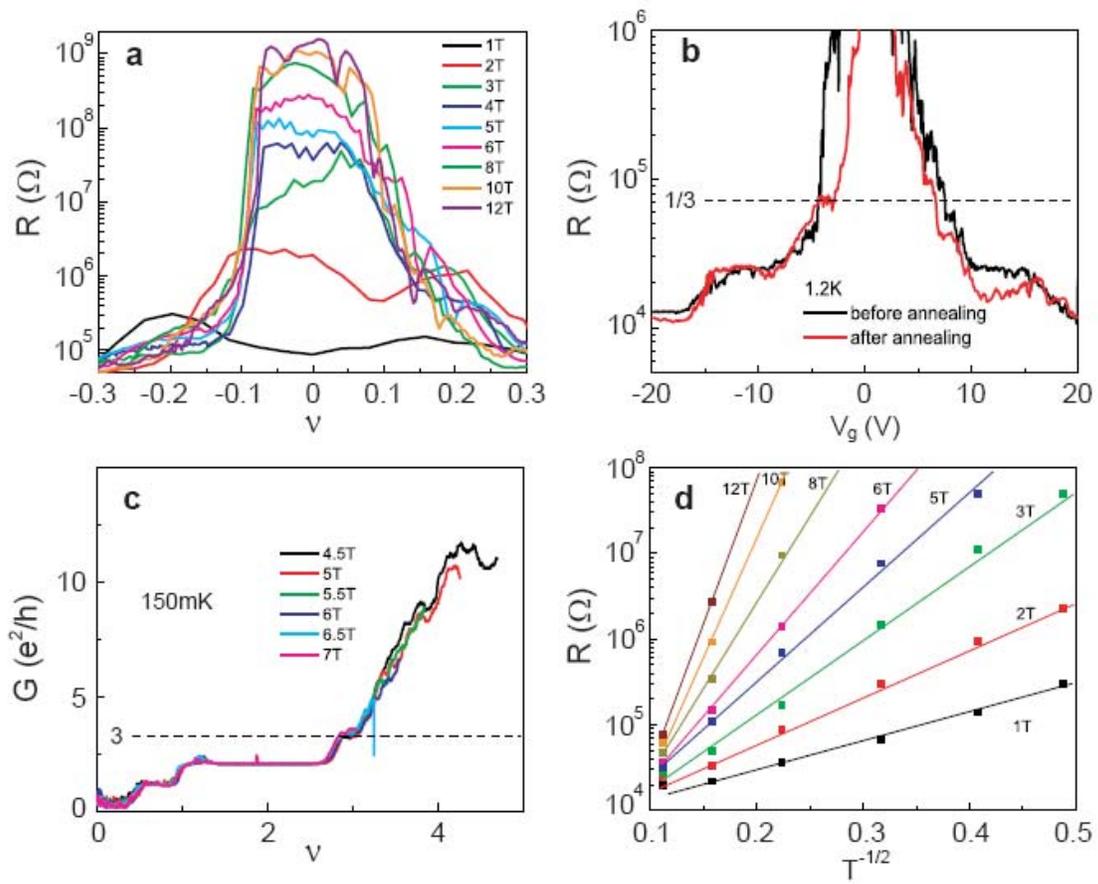

Figure 3